\documentclass[sigplan,10pt]{acmart}

\AtBeginDocument{%
  }

\usepackage{xspace}
\usepackage{xcolor}
\usepackage{makecell}
\usepackage{graphicx}
\usepackage{enumitem}
\setlist{leftmargin=1em, nosep}

\usepackage[frozencache=true,cacheignoresfilecontents=true,cachedir=_minted]{minted}
\setminted{
    fontsize=\fontsize{7.8}{8}\selectfont,
    frame=lines,
    framesep=4pt,
}
\usepackage{stfloats}

\usepackage{tikz}
\usetikzlibrary{intersections,arrows,chains,matrix,positioning,scopes,shapes,snakes}
\usetikzlibrary{positioning, arrows.meta, shapes.multipart}

\usepackage{caption}
\usepackage{subcaption}
\DeclareRobustCommand*\circleb[1]{\tikz[baseline=(char.base)]{
                \node[shape=circle,fill,inner sep=.5pt](char){\textcolor{white}{\small #1}};}}

\newcommand{\company}{ByteDance\xspace}
\newcommand{\sysname}{TensorHub\xspace}
\newcommand{\packagename}{tensorhub\xspace}
\newcommand{\minititle}[1]{\noindent\textbf{#1}\xspace}
\newcommand{\codeapi}[1]{\color{blue} \em #1}
\newcommand{\api}[1]{\texttt{#1}\xspace}

\makeatletter
\DeclareRobustCommand\onedot{\futurelet\@let@token\@onedot}
\def\@onedot{\ifx\@let@token.\else.\null\fi\xspace}

\def\eg{\emph{e.g}\onedot}

\def\ie{\emph{i.e}\onedot}

\copyrightyear{2026}
\acmYear{2026}
\setcopyright{cc}
\setcctype{by}
\acmConference[SOSP '26]{ACM SIGOPS 32nd Symposium on Operating Systems Principles}{September 29--October 02, 2026}{Prague, Czech Republic}
\acmBooktitle{ACM SIGOPS 32nd Symposium on Operating Systems Principles (SOSP '26), September 29--October 02, 2026, Prague, Czech Republic}
\acmDOI{10.1145/3830418.3843859}
\acmISBN{979-8-4007-2585-2/2026/09}

\begin{document}

\title[TensorHub: Scalable and Elastic Weight Transfer for LLM RL Training]{\sysname: Scalable and Elastic Weight Transfer\texorpdfstring{\\}{ }for LLM RL Training}

\author{
{\rm Chenhao Ye$^{\dagger}$, Huaizheng Zhang, Mingcong Han, Baoquan Zhong, Xiang Li, Qixiang Chen, \\
Xinyi Zhang, Weidong Zhang, Kaihua Jiang, Wang Zhang, He Sun, \\
Wencong Xiao$^{*}$, Andrea C. Arpaci-Dusseau$^{\dagger*}$, Remzi H. Arpaci-Dusseau$^{\dagger*}$} \\[0.75ex]
{ \rm \textit{ByteDance Seed} \qquad $^{\dagger}$\textit{University of Wisconsin--Madison} }
}

\makeatletter
\gdef\authors{Chenhao Ye, Huaizheng Zhang, Mingcong Han, Baoquan Zhong, Xiang Li, Qixiang Chen, Xinyi Zhang, Weidong Zhang, Kaihua Jiang, Wang Zhang, He Sun, Wencong Xiao, Andrea C. Arpaci-Dusseau, and Remzi H. Arpaci-Dusseau}
\g@addto@macro\@authornotes{\stepcounter{footnote}\footnotetext{Joint senior authors.}}
\makeatother
\renewcommand{\shortauthors}{Ye et al.}

\begin{abstract}

Modern LLM reinforcement learning (RL) workloads require a high-performance weight transfer system to scale training across heterogeneous compute resources.
However, efficiently transferring terabyte-scale model weights across thousands of GPUs remains challenging because the system must accommodate clusters that dynamically scale up and down while keeping coordination, data movement, and storage overhead low.

We introduce Reference-Oriented Storage (ROS), a new storage abstraction for RL weight transfer that exploits highly replicated model weights in place.
ROS presents the illusion that certain versions of the model weights are stored and can be fetched on demand.
Underneath, ROS does not physically store any copies of the weights; instead, it tracks the workers that hold these weights on GPUs for inference.
Upon request, ROS directly uses them to serve reads.
We build \sysname, a production-quality system that instantiates the ROS idea with topology-aware transfer, model-parallel consistency, and fault tolerance.
Evaluation shows that \sysname saturates RDMA bandwidth and adapts to three distinct rollout workloads with minimal engineering effort.
Specifically, \sysname reduces total GPU stall time by up to 6.7$\times$ for standalone rollouts, accelerates weight updates for elastic rollouts by up to 4.8$\times$, and cuts cross-datacenter rollout stall time by up to 19$\times$.
\sysname has been deployed in \company production to support cutting-edge RL training.

\end{abstract}




\maketitle
\pagestyle{empty}
\thispagestyle{empty}

\section{Introduction} \label{sec:introduction}

Reinforcement learning (RL) is central to building state-of-the-art large language models (LLMs)~\cite{hurst2024gpt,comanici2025gemini,Grok4,guo2025deepseek}.
With carefully designed rewards, RL aligns LLM responses with human preferences~\cite{RLHF,SafeRLHF} and improves reasoning~\cite{setlur2024rewarding,guo2025deepseek,comanici2025gemini}, coding~\cite{wei2025swe,liu2024rl}, and mathematical capabilities~\cite{DAPO,DeepSeekMath,setlur2024rl}.

Weight transfer plays a critical role in RL training.
RL operates as a continuous feedback loop between inference and training, with weight transfer closing the loop.
In each step, \emph{rollout} workers run inference to generate responses for reward scoring, and \emph{trainer} workers use the resulting data to update the model weights.
After each update, the new weights must be transferred to rollout workers for the next training step.
In addition, weight transfer is required for failure recovery, as failures are common at large scale.
Model weights are the heaviest state that a rollout worker must recover, and a restarted rollout worker cannot resume inference until it pulls the latest weights.

Production RL increasingly relies on diverse compute resources, posing new challenges for weight transfer.
For example, the \company production cluster has abundant spot GPU instances, but their frequent preemptions and restarts cause cluster membership churn, complicating their use.
Multi-datacenter deployments provide access to geographically distributed compute capacity~\cite{Multi-Datacenter-Training,StreamRL}, but slow inter-datacenter links make weight transfer a bottleneck.

Ideally, a weight transfer system should adapt to network topology and runtime load, accommodate cluster membership changes while masking failures, and avoid unnecessary control-flow coordination and data movement.
It should also require little additional storage or memory.

However, existing weight transfer approaches fail to satisfy these goals.
Collective communication (\eg, NCCL~\cite{NCCL}) delivers high throughput, but its reliance on static communication groups makes it brittle under failures and ill-suited for dynamic clusters with membership changes.
Point-to-point communication (\eg, UCX~\cite{UCX}) offers flexibility, yet suffers from network contention under fan-out.
Moreover, both approaches couple trainer and rollout control flow, forcing the RL framework to coordinate workers during weight transfer.

Another weight transfer approach is through distributed storage (\eg, parameter servers~\cite{ParameterServer,BytePS}, Ray object store~\cite{Ray}).
Unlike communication-based approaches, distributed storage offers a clean decoupling between trainers and rollouts: trainers \emph{push} new weights to storage, while rollouts \emph{pull} them on demand, all with minimal coordination; rollouts may scale up and down without trainers' awareness.
Yet, this push-then-pull pattern doubles data movement and incurs storage space or memory overhead, both of which are prohibitively expensive for TB-sized LLM weights.

We seek to retain the decoupling advantages of storage interfaces while eliminating the overhead of extra data movement and storage space.
This is made possible by an important observation: each version of the model weights is \emph{immutable} and \emph{highly replicated} for data-parallel inference.
Such redundancy naturally provides multiple equivalent data sources to serve read requests for that version.

Our key idea is to utilize highly replicated model weights \emph{in place} and provide access to them efficiently via a storage interface, called \textbf{Reference-Oriented Storage} (ROS).
ROS presents the illusion that certain versions of the model weights are stored and can be fetched on demand.
Underneath, ROS does not physically store any copies of the weights; instead, it tracks the workers that hold these weights on GPUs for inference. 
Upon request, ROS directly uses them to serve reads.
By avoiding explicit data ownership and extra copies, ROS combines the flexibility and simplicity of storage with the efficiency of direct transfer.

Building such a system is challenging due to \emph{integrity} and \emph{availability} requirements.
ROS addresses these challenges with two novel mechanisms: a \emph{mutability contract} and a \emph{retention protocol}.
The mutability contract disciplines the mutation semantics that enable safe weight buffer reuse.
The retention protocol defines which versions must be kept; in a corner case where a desired version may be lost, it offloads a copy to CPU memory to fulfill availability requirements.

We build \textbf{\sysname}, which instantiates the ROS idea in a production system with topology-aware transfer, model-parallel consistency, and fault tolerance.
It schedules transfers with topology awareness and leverages pipeline replication to scale throughput.
\sysname enforces a consistent view among workers within each model-parallel group to ensure correctness.
\sysname proactively detects failures and orchestrates recovery, enabling transparent operation under dynamic cluster scaling.

We evaluate \sysname on large-scale RL workloads spanning dozens to a thousand GPUs, elastic clusters with spot churn, and multi-datacenter topologies. 
In a 1024-GPU training task, \sysname reduces total GPU stall time by up to 6.7$\times$ compared to NCCL.
With an elastic cluster, \sysname's dynamic load balancing yields up to 4.8$\times$ faster updates than UCX.
In a cross-datacenter setting, \sysname cuts rollout GPU stall by up to 19$\times$ by avoiding redundant TCP transfers.

\sysname has been deployed in \company production to support cutting-edge RL training.
With its strong performance and extensive functionality, \sysname demonstrates a new way to build RL systems that operate effectively and 
efficiently at scale.

In summary, this paper makes the following contributions:

\begin{itemize}
\item Reference-Oriented Storage (ROS): an ownership-free storage abstraction specialized for model weights.
  
\item \sysname: a production-quality system that instantiates ROS with topology-aware transfer, model-parallel consistency, and fault tolerance.
  
\item Three case studies on diverse rollout workloads to demonstrate \sysname's performance and functionality benefits over other production baselines.
\end{itemize}

\section{Background and Motivation} \label{sec:background}

In modern LLM reinforcement learning (RL) workloads, weight transfer is essential for fully leveraging heterogeneous compute resources. 
This section covers the background of RL workloads, the goals of a weight transfer system, and the limitations of existing approaches.

\begin{figure}[t]
    \centering
    \includegraphics[width=\linewidth]{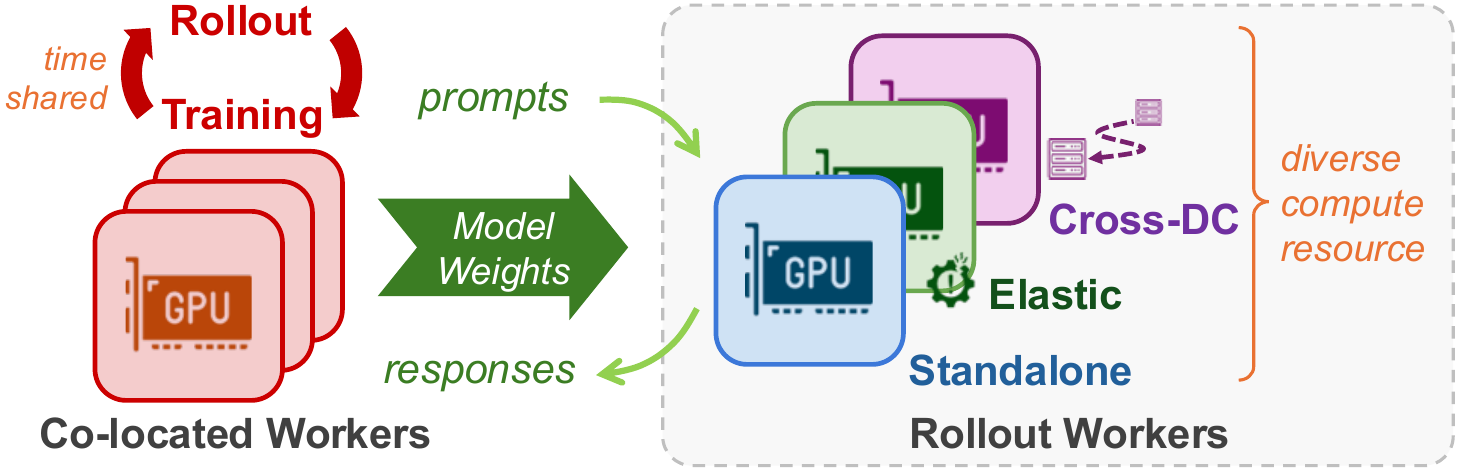}
    \caption{\textbf{RL Workload with Diverse Rollout Patterns.} 
    \it
    Rollout workers stream in prompts and model weights and stream out responses. While prompts and responses are lightweight, the model weights can be several terabytes in size, necessitating an efficient transfer system.
    }
    \label{fig:background}
\end{figure}

\subsection{RL Workload Background} \label{sec:background:rl}

A typical RL algorithm~\cite{PPO,DeepSeekMath} proceeds through three stages: (1)~\emph{rollout}, which runs inference and generates responses for given prompts; (2)~\emph{scoring}, which judges the responses with a reward module (\eg, human ratings~\cite{PreferenceRL,RLHF,HH_RLHF}, a reward model~\cite{PreferenceRL,RLHF}, or rule-based rewards~\cite{DeepSeekMath}); and (3)~\emph{training}, which computes gradients and updates the model weights.
The updated weights are used in the next rollout, forming a continuous feedback loop.

RL frameworks implement the algorithms by mapping rollout and training onto GPU workers.
Figure~\ref{fig:background} illustrates the rollout and training patterns used by representative RL frameworks.
In a basic \emph{co-located} architecture, rollout and training time-share the same set of physical GPUs.
We refer to these co-located workers as \emph{trainers}.
This architecture is widely adopted by veRL~\cite{HybridFlow}, ReaL~\cite{ReaL}, and RLHFuse~\cite{RLHFuse} for its simplicity and high resource utilization.

However, rollout time can account for more than 90\% of training time~\cite{OpenRLHF,RhymeRL,HybridFlow}, thereby becoming the bottleneck.
This motivates many new patterns to scale out rollout by utilizing additional GPUs:

\begin{itemize}

\item \emph{Standalone rollouts} are GPU workers dedicated to rollout tasks. This design is widely adopted by many frameworks~\cite{AReaL,LlamaRL,Laminar,OpenRLHF}.
  
\item \emph{Elastic rollouts} run on spot GPU instances that can be preempted or restarted at any time. Their lower price enables greater cost efficiency. RLBoost~\cite{RLBoost} reports up to 1.97$\times$ higher training throughput with this approach.

\item \emph{Cross-datacenter rollouts} run on GPUs located in a datacenter separate from the trainers. They enable hyperscale training~\cite{Multi-Datacenter-Training} and leverage heterogeneous resource availability~\cite{StreamRL}.
Because GPUs are often procured and deployed in different batches, inference-optimized GPUs may not be available in the trainers' datacenter~\cite{StreamRL}.
\end{itemize}

With rollouts running on dedicated GPUs, model weights must be transferred from trainers after each training step, typically every few to tens of minutes.
At scale, each update disseminates terabyte-sized model weights to thousands of GPUs.
This transfer lies on the critical path: rollout workers cannot perform inference using partially updated weights, leaving their GPUs idle until the weight transfer is complete.

In addition, rollout workers are organized into model-parallel groups, each holding a full model copy sharded across its workers.
These groups are loosely coupled with one another, progress at different paces as they generate variable-length outputs, and become ready for new weights at different times.
The weight transfer system must serve these transfers dynamically across groups.

\begin{table*}
    \centering\small
    \begin{tabular}{c|c|ccccc}
    \hline\hline
        \textbf{Approach}  & \textbf{Example} & \makecell{Topology\\Awareness} & Elasticity & \makecell{Minimal\\ Coordination} & \makecell{Minimum \\Data Movement} & \makecell{No Extra\\Space} \\
    \hline
        Collective Communication & NCCL~\cite{NCCL}   & \checkmark  & & & \checkmark & \checkmark\\
        Point-to-Point Communication & UCX~\cite{UCX} & & \checkmark &  & \checkmark  & \checkmark\\
        Distributed Storage & PS~\cite{ParameterServer}, Ray object store~\cite{Ray} & \checkmark & \checkmark & \checkmark &   &  \\
    \hline
        \multicolumn{2}{c|}{\sysname} & \checkmark & \checkmark & \checkmark & \checkmark & \checkmark \\
    \hline\hline
    \end{tabular}
    \caption{\textbf{Comparison of Different Weight Transfer Approaches.}
    }
    \label{tab:existing}
\end{table*}

\subsection{Weight Transfer System Goals} \label{sec:background:goal}

To serve these diverse rollout patterns, an ideal weight transfer system must satisfy five goals: 

\minititle{Goal 1: Topology Awareness.}
To fully utilize network bandwidth, the weight transfer system must incorporate network topology and runtime load knowledge.
For cross-datacenter training, it must minimize the network traffic over slow inter-datacenter links.
Within the same datacenter, it should avoid single-node bandwidth congestion.

\minititle{Goal 2: Elasticity.}
The weight transfer system must support a dynamic cluster that scales up and down while transparently masking failures.
If a worker fails or is preempted, the system must safely evict it without disrupting healthy workers.
If a worker restarts or joins the cluster, the system must promptly serve weights on demand.

\minititle{Goal 3: Minimal Coordination.}
Weight transfer should minimize control-flow coordination among workers.
Coordination introduces temporal coupling, where workers must stop computation to cooperate.
As a result, it not only complicates the programming model but also leaves GPUs idle while waiting and amplifies stragglers at scale.

\minititle{Goal 4: Minimum Data Movement.}
The optimal data transfer must occur directly between the source and destination GPUs, with no intermediate hop.
An extra hop implies additional network or PCIe costs, hurting overall efficiency.

\minititle{Goal 5: No Extra Storage or Memory Space.}
As LLM weights can be several terabytes, storing extra copies requires massive storage or memory space. Such extra resource requirements should be minimized for cost efficiency.

\subsection{Limitations of Existing Approaches} \label{sec:background:existing}

Existing weight transfer approaches fall into three main categories: \emph{collective communication}, \emph{point-to-point communication}, and \emph{distributed storage}. Table~\ref{tab:existing} summarizes how well each meets the above goals.

\minititle{Collective Communication.}
The NVIDIA Collective Communication Library (NCCL)~\cite{NCCL} is widely used for weight transfer, adopted by veRL~\cite{HybridFlow}, OpenRLHF~\cite{OpenRLHF}, AReaL~\cite{AReaL}, vLLM~\cite{vLLM}, SGLang~\cite{SGLang}, Slime~\cite{slime_github}, and others.
In these frameworks, trainers broadcast weights to rollouts via NCCL collectives.
NCCL detects the hardware topology and constructs ring or tree schedules that saturate RDMA bandwidth.

However, collective communication fails on the elasticity and minimal coordination goals.
First, collective communication requires a pre-defined communication group that cannot add or remove nodes afterward. This prevents it from being applied to a dynamic cluster where nodes may fail, be preempted, or be restarted; any failure will cause the entire cluster to crash.
Second, broadcast requires all nodes to cooperate, which in practice translates to a global barrier in RL frameworks that interrupts all workers and does not resume execution until all transfers have finished~\cite{HybridFlow,AReaL,OpenRLHF}.
This amplifies stragglers, which are common in RL workloads. For example, a rollout may delay responding to an interrupt while generating a long sequence, or network contention may slow a subset of nodes during transfer.

\minititle{Point-to-Point Communication.}
Point-to-point communication (\eg, UCX~\cite{UCX}) enables flexible data transfer between workers without a communication group, and thus naturally supports dynamic clusters.
In an ideal case, such direct transfer can saturate the RDMA bandwidth between two endpoints, delivering high throughput.

However, point-to-point communication is less commonly used in RL because it has no global view of the runtime topology and therefore cannot balance load effectively.
Senders serve requests independently, making their outbound bandwidth the bottleneck under fan-out.
For example, when multiple rollouts pull weights from a trainer, they contend for the trainer's uplink bandwidth, degrading performance.
Moreover, point-to-point communication still requires coordination at the framework level. When a trainer issues a \texttt{send()}, the corresponding rollout must be interrupted to invoke \texttt{recv()}; the \texttt{send()} blocks until the receiver cooperates.

\minititle{Distributed Storage.}
Distributed storage systems, \eg, parameter servers (PS)~\cite{ParameterServer,BytePS} and the Ray object store~\cite{Ray}, provide a compelling \emph{decoupled} interface for weight transfer: trainers \emph{push} updated weights to storage, and rollouts \emph{pull} weights on demand. 
By replicating weights across storage nodes, such systems can exploit topology and locality, serving requests from nearby or less-loaded replicas. 
Moreover, this decoupling also naturally supports elasticity, since rollouts can join, leave, or restart independently without trainers' awareness. 
Finally, trainers can proceed immediately after pushing weights, while rollouts fetch weights on demand, minimizing coordination between them.

Despite these compelling properties, storage has not been adopted as the primary weight transfer mechanism in major LLM RL frameworks due to poor performance.
For parameter servers~\cite{ParameterServer,BytePS}, the push-then-pull workflow amplifies network traffic, and the storage server's network bandwidth becomes a potential bottleneck at scale.
The Ray object store~\cite{Ray} mitigates the network bottleneck by co-locating storage with workers, but still incurs extra data movement, requiring GPU--CPU copies and (de)serialization before transferring and reconstructing weights.
For example, transferring 40~GB of data between two 8-GPU machines via the Ray object store takes 32~s in our measurement, whereas GPU-direct RDMA completes the same transfer in $\sim$0.2~s. Such orders-of-magnitude performance losses can hardly be justified by functionality benefits.
Lastly, storing TB-sized LLM weights is too expensive in either remote or co-located storage.
In our experiment, the Ray object store commonly crashes due to out-of-memory errors when transferring more than 300~GB of data at once.

More fundamentally, data \emph{ownership} is the root cause of storage's extra movement and space overhead.
Each write operation copies weights into storage-managed space (either a remote server or co-located CPU memory), allowing the system to manage data independently of workers. This ownership enables trainer-rollout decoupling but also introduces additional serialization, copying, network traffic, and storage overhead that ultimately degrade performance.

\minititle{Summary.}
Existing approaches present a trade-off. Collective communication achieves high throughput with topology-aware transfer, but requires rigid membership and global coordination. Point-to-point communication supports dynamic clusters and direct transfer, but struggles with load balancing and still incurs coordination at runtime. Distributed storage offers clean decoupling and flexibility, but pays the price of additional data movement and storage overhead due to full data ownership. As a result, none of these approaches satisfies all five goals---topology awareness, elasticity, minimal coordination, minimum data movement, and no extra storage or memory space.

\section{Reference-Oriented Storage} \label{sec:idea}

Distributed storage offers a clean and flexible interface for weight transfer, but its reliance on data ownership introduces fundamental inefficiencies.
To address this limitation, we present \emph{Reference-Oriented Storage} (ROS), a new storage abstraction that does not own data. 

ROS is enabled by three important properties of model weights in RL.
First, weights are \emph{highly replicated} due to data parallelism. This redundancy means there are multiple symmetric sources available for transfer.
Second, weights used in inference are \emph{immutable}. As a result, reading them from a remote worker's GPU memory is safe and requires minimal coordination.
Third, weight versions are \emph{short-lived}.
Training constantly generates new versions, and only the most recent few are relevant, so long-term durability is unnecessary.

ROS directly repurposes those model weights \emph{in place} to provide the illusion that certain versions of the weights are stored and can be fetched on demand.
Under the hood, ROS does not physically store any copies of the weights.
Instead, it tracks the workers with these weights on GPUs for inference; upon request, ROS directly uses them to serve reads.
By avoiding explicit data ownership and extra copies, ROS combines the flexibility of storage with the efficiency of direct transfer.

This section covers the elemental primitives required for a minimal ROS system.
We first present an overview of ROS and then elaborate on two key mechanisms required for integrity and availability.
A more sophisticated system with rich features will be covered in \S\ref{sec:design}.

\subsection{ROS Overview} \label{sec:idea:semantics}

\begin{figure}[t]
    \centering
    \includegraphics[width=\linewidth]{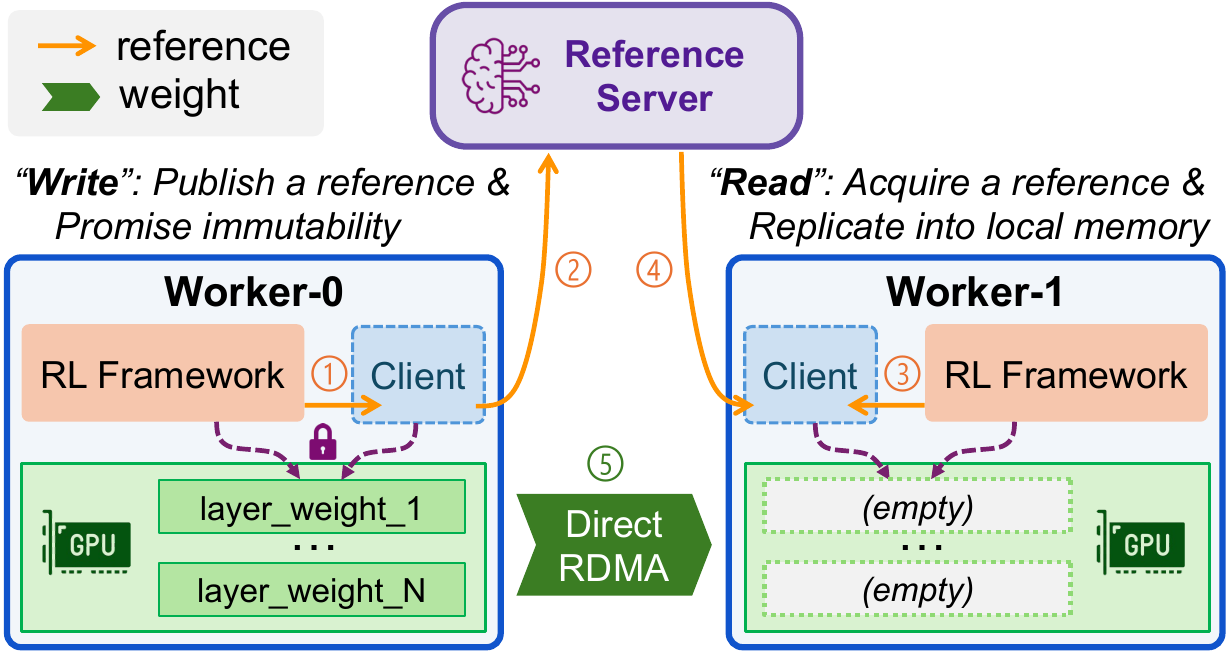}
    \caption{\textbf{Reference-Oriented Storage Workflow.}
    \it
    The server only operates on lightweight references. The bulk weight transfer is directly between clients.
    }
    \label{fig:ros_arch}
\end{figure}

ROS consists of a centralized reference server and a client library imported by each worker.
Figure~\ref{fig:ros_arch} illustrates an example where Worker-0 holds a version of model weights in GPU memory for inference, and Worker-1 needs that version.

ROS exposes \api{publish} and \api{replicate} primitives for write and read operations.
\circleb{1} To write weights into ROS, Worker-0 calls \api{publish} and passes a reference to its in-GPU weights to its client.
\circleb{2} The client forwards this reference to the reference server.
\circleb{3} To read weights, Worker-1 invokes \api{replicate} with a reference to its local weight buffers.
\circleb{4} Its client queries the server, which returns a source reference pointing to Worker-0's GPU memory.
\circleb{5} Worker-1's client then issues RDMA to read the weights directly from Worker-0 into its local buffers.
Upon completion, Worker-1's copy becomes another replica of the model weights. Future replication requests can then be served from either worker.

Critically, these data flows require no additional data movement or storage overhead.
Unlike traditional distributed storage, the ROS server never stores or transfers weight data. It only operates on the lightweight references.
The bulk data transfer is directly between the source and destination memory without intermediate hops or staging space.

ROS treats all copies of the same weights as symmetric replicas, whether created via \api{publish} or \api{replicate}.
A rollout thus need not fetch weights from a trainer; it can retrieve them from any peer, enabling efficient peer-to-peer data dissemination.
Upon failure, a worker can restart and recover from any surviving replica, making the cluster self-healing.

\minititle{Challenges.}
Since ROS only \emph{references} weights but does not \emph{own} them, it must address two challenges to be functional.

\begin{itemize}
\item \emph{Integrity}: Weights require the flexibility for occasional mutations (\eg, buffer reuse for a new version). ROS must define clear mutation semantics to avoid data corruption.

\item \emph{Availability}: Because ROS does not own the data, a weight version disappears if no worker holds a copy. The system must define which weights should remain accessible and introduce protocols to preserve that.
\end{itemize}

\noindent ROS introduces two key mechanisms to address the challenges: the \emph{mutability contract} and the \emph{retention protocol}.

\subsection{Integrity via Mutability Contract} \label{sec:idea:mutability}

ROS relies on weight immutability for safe sharing, but RL training requires the flexibility for mutability.
For example, as weights are too large to be duplicated on GPUs, a rollout may want to reuse the weight buffers in place when pulling the new weights. Similarly, a trainer must be able to mutate the weights to produce a new version.

ROS addresses this demand with the \api{unpublish} primitive and the \emph{mutability contract}.
When holding a replica, a worker commits not to mutate the weight data.
When mutation is required, the worker must call \api{unpublish} to explicitly revoke the immutability commitment.
Upon request, the reference server ensures this worker's weights are no longer visible to others; if there is an ongoing transfer from this worker, the server waits for the transfer to drain and only acknowledges unpublish success afterward. This ensures every concurrent transfer still receives uncorrupted data.

Fortunately, adhering to the mutability contract is easy in practice.
An RL worker's execution flow is typically organized into stages, in which the immutability boundary is straightforward to identify.
In addition, RL frameworks are often implemented in Python, whose memory safety avoids many dangling-pointer bugs.

\subsection{Availability via Retention Protocol} \label{sec:idea:retention}

ROS stores no copies, so if all workers unpublish a version, that version is unavailable.
Such behavior is appropriate for overly stale weights, as they are not needed by any workers and thus safe to drop.
However, losing the latest (few) versions of the weights is unacceptable.
This scenario can occur in a corner case if all trainers unpublish the latest version before it is replicated by any rollout.
To avoid such unavailability, the system requires clear semantics for ``what data must be available'' and a protocol to enforce that.

ROS fulfills this requirement with the \emph{retention protocol}.
ROS exposes \api{retain} semantics under which a worker can declare the versions to keep available (\eg, the latest $k$ versions).
When processing \api{unpublish} requests, the server checks if the requesting replica is the last copy of a retained version; if so, it notifies the client to offload a copy of the weights to CPU memory and publish that as an \emph{offload replica}.

Offloading is a safeguard for corner cases and is rarely triggered in practice.
Rollouts typically see and replicate the new version promptly; as long as one copy is replicated, trainers can unpublish without triggering offload.
When offloading does occur, only a single copy across the entire cluster is kept, minimizing memory overhead. 
Measurements show that offloading over the PCIe bus is fast ($\sim$48~GB/s in our setup), resulting in a minor delay for unpublish (<2~s even when offloading the entire GPU's weights).
The offload replica is automatically released once it has been replicated by another worker (and is thus no longer the last copy) or once a newer version's publish makes it no longer retained.

The elegance of the retention protocol is that it provides similar availability to traditional ownership-based storage, while eliminating the ownership overhead in common cases.

\subsection{ROS Summary}

ROS achieves Goals 3, 4, and 5 (defined in \S\ref{sec:background:goal}) by design.
By providing a storage interface, ROS enables rollouts to pull weights on demand without trainers' explicit coordination (Goal 3). 
By not taking ownership of weight data, ROS avoids the overhead of data movement (Goal 4) and extra storage space (Goal 5) in the common case.

ROS also establishes a good foundation for Goals 1~and~2.
In principle, its centralized server architecture offers a global view of network topology and runtime load, making topology-aware transfer easier to implement (Goal 1).
Elasticity can be achieved as long as the ROS system itself is fault-tolerant (Goal 2).
In summary, both goals fit into ROS's design but require a well-optimized implementation.

\section{\sysname Design and Implementation} \label{sec:design}

We build \sysname, a production-quality system that instantiates the ROS idea with rich features required by real-world use cases.
As LLM sizes increase, model parallelism is required, with each GPU worker holding only one shard of weights.
Therefore, sharding must be natively incorporated in addition to the previous ROS requirements.

Overall, a production-quality system must satisfy four practical requirements.
First, it must provide concise and expressive interfaces for easy integration into existing RL frameworks.
Second, for efficient transfer at scale, the system must schedule transfer based on the network topology and runtime load.
Third, for correctness, the system must guarantee that workers in the same model-parallel group see a consistent view of the system state, preventing control-flow divergence.
Finally, to support dynamic cluster scaling up and down, it must gracefully handle failures.

This section first describes the naming scheme (\S\ref{sec:design:naming}) and APIs (\S\ref{sec:design:api}) and then introduces topology-aware transfer (\S\ref{sec:design:replication}), model-parallel consistency (\S\ref{sec:design:consistency}), fault tolerance (\S\ref{sec:design:fault}), and implementation (\S\ref{sec:design:implementation}).

\subsection{Naming Scheme} \label{sec:design:naming}

\begin{table*}
    \centering
    \small
    \begin{tabular}{p{0.21\linewidth}p{0.75\linewidth}}
    \hline
        \textbf{Name} & \textbf{Description}\\
    \hline
        \api{\textbf{open}($\ldots$) -> ShardHandle} & 
        Acquire a shard handle. Optionally, declare the desired versions to retain for the retention protocol. \\
    \hline
        \api{\textbf{register}(named\_tensors)} & 
        Register weight tensors. \\
    \hline
        \api{\textbf{unregister}()} & 
        Unregister weight tensors. \\
    \hline
        \api{\textbf{publish}(version)} & 
        Publish a reference to the registered tensors under the given version. \\
    \hline
        \api{\textbf{unpublish}()} & 
        Revoke the previously published reference. \\
    \hline
        \api{\textbf{replicate}(version)} & 
        Replicate the given version of weights into registered tensors. \emph{Supports relative versions.} \\
    \hline
        \api{\textbf{update}(version) -> bool} & 
        Atomically check whether the given version is available and differs from the current one; if so, unpublish the current version and replicate the new one; return true if updated.
        \emph{Supports relative versions.} \\
    \hline
        \api{\textbf{list}() -> Dict} & 
        Return a dictionary mapping each available version to its set of replica names. \\
    \hline
        \api{\textbf{wait}(predicate)} & 
        Wait until \api{predicate(list())} evaluates to true. \\
    \hline
        \api{\textbf{close}()} & 
        Destroy the handle; automatically unpublish and unregister if applicable. \\
    \hline
    \end{tabular}
    \caption{\textbf{\sysname APIs.} \it \api{open()} returns a handle, and the rest of the APIs are methods of this handle. }
    \label{tab:api}
\end{table*}

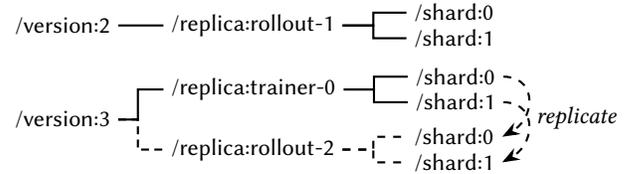
\begin{figure}[t]
\centering
\small
\tikzset{
  every node/.style={font=\sffamily},
  fork_out/.style={to path={(\tikztostart) -- ++(#1,0) |- (\tikztotarget) \tikztonodes}},
  solid_arrow/.style={thick},
  dashed_arrow/.style={thick, densely dashed},
  hollow_arrow/.style={thick, dashed}
}

\begin{tikzpicture}[
  node distance=0.4cm and 0.6cm
]

\node (v2) {/version:2};
\node (rep1) [right=of v2] {/replica:rollout-1};
\node (shard1_0) [right=of rep1, yshift=0.17cm, xshift=0.25cm] {/shard:0};
\node (shard1_1) [right=of rep1, yshift=-0.17cm, xshift=0.25cm] {/shard:1};

\draw[solid_arrow] (v2) -- (rep1);
\draw[solid_arrow] (rep1) to[fork_out=1.6cm] (shard1_0.west);
\draw[solid_arrow] (rep1) to[fork_out=1.6cm] (shard1_1.west);

\node (v3) [below=0.8cm of v2] {/version:3};

\node (rep2) [right=of v3, yshift=0.4cm] {/replica:trainer-0};
\node (shard2_0) [right=of rep2, yshift=0.17cm, xshift=0.25cm] {/shard:0};
\node (shard2_1) [right=of rep2, yshift=-0.17cm, xshift=0.25cm] {/shard:1};

\node (rep3) [right=of v3, yshift=-0.4cm] {/replica:rollout-2};
\node (shard3_0) [right=of rep3, yshift=0.17cm, xshift=0.25cm] {/shard:0};
\node (shard3_1) [right=of rep3, yshift=-0.17cm, xshift=0.25cm] {/shard:1};

\draw[solid_arrow] (v3) to[fork_out=1cm] (rep2.west);
\draw[dashed_arrow] (v3) to[fork_out=1cm] (rep3.west);

\draw[solid_arrow] (rep2) to[fork_out=1.6cm] (shard2_0.west);
\draw[solid_arrow] (rep2) to[fork_out=1.6cm] (shard2_1.west);

\draw[hollow_arrow] (rep3) to[fork_out=1.6cm] (shard3_0.west);
\draw[hollow_arrow] (rep3) to[fork_out=1.6cm] (shard3_1.west);

\draw[dashed, thick, -Stealth]
  (shard2_0.east) .. controls +(0.45,0) and +(0.45,0.25) ..
  (shard3_0.east);

\draw[dashed, thick, -Stealth]
  (shard2_1.east) .. controls +(0.45,0) and +(0.45,0.25) ..
  node[right, pos=0.4] {\rm \emph{replicate}}
  (shard3_1.east);

\end{tikzpicture}
\caption{\textbf{\sysname Naming Scheme.}
\it
Rollout-2 fetches version 3 by replicating a copy; after doing so, it also becomes a replica capable of serving subsequent replication requests.
}
\label{fig:data_scheme}
\end{figure}

We begin with a hierarchical naming scheme to establish a precise vocabulary.
In \sysname, each model forms an independent domain, managed by a single reference server.
Each model evolves over \emph{versions}, each produced by one step of training.
A version consists of one or more \emph{replicas}, each representing a full copy owned by a model-parallel worker group. 
Finally, each replica is divided into \emph{shards}, with each shard owned by a single worker.
Figure~\ref{fig:data_scheme} illustrates an example with three replicas, each composed of two shards.

Each version is uniquely identified by an integer, called the \emph{absolute} version.
\sysname also supports \emph{relative} versions, where the strings ``latest'' or ``latest-$k$'' are resolved to the latest version or the latest minus $k$, respectively.
Relative versions are useful because RL typically only cares about freshness relative to the latest one. 
Off-by-$k$-step RL training workloads~\cite{April,LUFFY,AReaL,Laminar,LlamaRL} may require multiple versions to coexist in the system, where ``latest-$k$'' is useful.

\subsection{\sysname APIs} \label{sec:design:api}

\sysname APIs are summarized in Table~\ref{tab:api}. 
As the same set of tensor buffers is commonly used across versions, \sysname exposes \texttt{ShardHandle} to operate on them.

\begin{itemize}
\item \minititle{Open.}  
A worker begins by calling \api{\packagename.open()} to acquire a handle for a shard within a replica. 
If it wishes to retain certain versions, it specifies them upon open.

\item \minititle{Registration.} 
Before publishing or replicating, a worker must register weight tensors with the handle; before deallocating these tensors, it must unregister.
These APIs support both CPU and GPU tensors, offering flexibility in memory management.
For example, trainers in fully disaggregated architectures~\cite{LlamaRL,StreamRL} execute training in a tight loop without co-located rollout and thus constantly mutate their on-GPU weights; they can publish weights from CPU memory.

\item \minititle{Publish.}
To make its registered tensors visible under version $v$, the worker calls \api{publish(v)} with an immutability commitment.
When the worker desires mutation, it must invoke \api{unpublish()} first. These two calls form the core mutability discipline for reference safety.

\item \minititle{Replicate.}
A worker obtains weights by naming the desired version (absolute or relative). \api{replicate(v)} materializes version $v$ into the handle's registered tensors, blocking if necessary until the version exists.   

\item \minititle{Update.}
With a version held, a worker can call \api{update(v)} to atomically transition to a newer version if available. 
Atomicity avoids races between a version check and replication.
This API is useful for a rollout to poll whether a new version of the weights is available; if not, it continues inference with the current weights.

\item \minititle{Query.}
\sysname provides lightweight introspection through \api{list()} and \api{wait(predicate)}, which expose global metadata (such as available versions or replicas) for implementing custom policies. For example, rollouts can discard excessively stale responses, or trainers can wait for lagging rollouts to catch up~\cite{April,LUFFY,AReaL,Laminar,LlamaRL}.

\item \minititle{Close.}
When a worker finishes using the handle, it calls \api{close()} to release resources and clean up state.
\end{itemize}

\begin{figure}[t]
    \centering
    \begin{subfigure}[b]{0.47\linewidth}
         \input{code/colocated_rollout}
         \caption{Trainer.}
         \label{fig:code:colocated}
    \end{subfigure}
    \hfill
    \begin{subfigure}[b]{0.47\linewidth}
         \input{code/dedicated_rollout}
         \caption{Standalone rollout.}
         \label{fig:code:dedicated}
    \end{subfigure}
    \caption{\textbf{\sysname Examples.}}
    \label{fig:code}
\end{figure}

\minititle{Example.}
Figure~\ref{fig:code:colocated} shows the workflow of a trainer that runs both training and rollout tasks.
At the beginning of each step, it publishes the current weights and starts rollout.
After completing a configured amount of rollout work, it unpublishes the weights and switches to training.
Once training completes, it converts the updated weights to an inference-ready format (\eg, resharding and quantization), publishes the new version, and begins the next rollout step.

Figure~\ref{fig:code:dedicated} shows an example of a standalone rollout.
Upon initialization, it calls \api{replicate()} to fetch a copy of the weights before starting inference.
Then, after finishing a batch of inference or receiving a control message, it calls \api{update()} to fetch a new version if available; otherwise, it continues inference with the existing weights.

%%%%%%%%%%%%%%%%%%%%%%%%%%%%%%%%%%%%%%%%%%%%%%%%%%%%%%%%%%%%%%%%%%%%%%%%%%%%%%%%

\subsection{Topology-Aware Replication at Scale} \label{sec:design:replication}

In RL training, every newly published version often needs to be fetched by multiple rollouts. A system that funnels all replication through a single node will collapse under network contention. 
\sysname is therefore architected with a centralized reference server to schedule transfer based on network topology and runtime load, while the clients remain decentralized for scalable data movement.

\subsubsection{Load-Balanced Scheduling at the Server}

Efficient transfer begins with choosing the right source for each request. 
\sysname centralizes scheduling on the server, exploiting its global view of the system for informed decisions.

The server always prioritizes a source replica located in the same datacenter.
Within each datacenter, the server maintains a reference count for each replica---representing the number of replication requests it is currently serving---and always selects the least-loaded replica.
Under pipeline replication (\S\ref{sec:design:architecture:pipeline}), each replica serves at most one request at a time, barring failures.  
This load-balancing policy also keeps unpublish latency low: if a source replica requests to unpublish, it needs to wait for in-flight replication to drain, which is brief and bounded by a single request.
If no source replica is available within the same datacenter, the server falls back to cross-datacenter replication (\S\ref{sec:design:crossdc}).

In practice, a single server is sufficient to scale a training task to thousands of workers (more measurements in \S\ref{sec:evaluation:microbench:server}).
As future work, if the server becomes a scalability bottleneck at even larger scales, \sysname could partition workers into sub-clusters, each managed independently by a server.
These servers would only need to lazily synchronize on critical events (\eg, a new version is published).

\subsubsection{High-Performance Transfer between Clients}

\sysname's data-plane efficiency relies on a high-performance transfer method between clients.
To achieve this, each client embeds a hardware-affinity-aware transfer engine that uniformly supports RDMA and TCP transfers over both GPU and CPU memory.
More specifically, the transfer engine offers three transfer modes, customized for workloads and available transport methods:

\begin{itemize}
\item \emph{RDMA Direct}:
In workloads where tensors remain allocated for long periods, the transfer engine registers tensor buffers directly with the RDMA NIC. Remote clients can then issue one-sided RDMA reads to pull data directly. This zero-copy path bypasses CPUs entirely and achieves the best performance.

\item \emph{RDMA Copy}: When users frequently reallocate tensors, repeated RDMA NIC registration is too costly. In this case, the transfer engine employs background threads to serve remote requests by copying the tensor data slice by slice into pre-registered buffers and then issuing RDMA writes. 
This approach eliminates re-registration overhead while still preserving RDMA's high throughput.

\item \emph{TCP}: In environments where RDMA is unavailable (\eg, across datacenters), the transfer engine uses TCP.
\end{itemize}

The transfer engine is hardware-affinity-aware and optimized for modern multi-GPU, multi-NIC servers. For GPU-resident tensors, the engine selects the NIC closest to the tensor's GPU; in RDMA copy mode, it uses the pre-registered RDMA regions on the same GPU devices.

\minititle{Tiny-Tensor Optimization.}
LLMs contain a large number of tiny tensors, which are inefficient to register and transfer.
To address this, \sysname compacts all tiny tensors (<2~MB) into contiguous buffers; only these compacted buffers are registered and transferred.
During replication, the receiver unpacks them into their corresponding slices. 

This optimization reduces registration costs and improves bandwidth utilization, at the cost of very minor memory overhead.
For example, a 36B model has approximately five hundred tensors, half of which are tiny. When compacted, they require only about 3~MB of extra memory in total, which is negligible for a 19~GB shard.

\subsubsection{Pipeline Replication}

\begin{figure}[t]
    \centering
    \includegraphics[width=\linewidth]{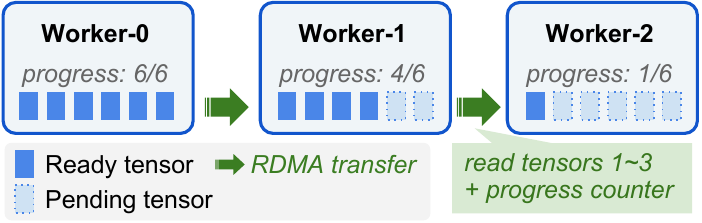}
    \caption{\textbf{Pipeline Replication.} 
    \it
    Worker-0 is the only source, while both Worker-1 and Worker-2 are requesting data.
    To scale throughput, \sysname schedules a pipeline where Worker-2 reads partially replicated data on Worker-1.
    }
    \label{fig:pipeline}
\end{figure}

\label{sec:design:architecture:pipeline}

Pipeline replication enables \sysname to scale throughput beyond what a single source could provide. A key challenge in RL training is that large bursts of rollouts may simultaneously request the latest version, creating extreme fan-out. Pipeline replication allows partially replicated workers to begin serving others immediately, turning replication into an expanding pipeline rather than a single-rooted broadcast (as shown in Figure~\ref{fig:pipeline}). Pipeline replication builds on the observation that most RDMA NICs are full-duplex, and therefore, when a NIC is receiving data, its uplink is idle and can serve data.

Pipeline replication uses the metadata and transfer mechanisms described below. Each replicating worker tracks a \emph{progress counter} that indicates how many tensors of the target version it has received locally. When a new worker requests a version, the reference server selects the least-loaded source, which can be either fully replicated or still in progress. If the source is only partially replicated, the requester reads the source's progress counter and then
fetches the prefix that is available. It then re-reads the progress counter and repeats until all data is received.

Pipeline replication transforms the topology from a fixed N-way fan-out into a bandwidth-amplifying DAG, where each replicating worker increases the system's bandwidth. Because transfers are decentralized and occur directly between peers, throughput grows with the number of active clients, and latency remains low even under heavy demand.

\subsubsection{Cross-Datacenter Replication} \label{sec:design:crossdc}

\sysname's replication model generalizes to clusters spanning multiple datacenters, with a two-stage replication pattern.

When a new version is published, the first replica located in a different datacenter must fetch the data over TCP; we refer to this replica as the \emph{seeding replica}. Once seeding completes, subsequent replicas in that datacenter can fetch the data over RDMA, enabling full pipeline replication locally without repeatedly invoking cross-datacenter transfer.

TCP-based seeding has substantially lower bandwidth than RDMA and can therefore stall pipeline replication if other replicas immediately try to update and fetch from the partially filled seeding replica. To avoid this stall, \sysname augments the \api{update()} API with \emph{smart skipping}: if a polling replica observes that a seeding replica is actively fetching over TCP, it treats the new version as temporarily unavailable and retries later, allowing seeding to complete before datacenter-local replication begins. This avoids serializing the entire pipeline behind slow cross-datacenter transfer.

To further reduce GPU stall time for seeding replicas, \sysname offers an \emph{offload seeding} option. When seeding is required, the client library creates a temporary offload buffer in CPU memory and initiates cross-datacenter transfer into this buffer in the background.
The \api{update()} call returns immediately, and subsequent polling prioritizes consuming the offloaded copy if completed. This approach trades a bounded amount of CPU memory for reduced GPU stall time by creating at most one offload seeding replica per datacenter.

%%%%%%%%%%%%%%%%%%%%%%%%%%%%%%%%%%%%%%%%%%%%%%%%%%%%%%%%%%%%%%%%%%%%%%%%%%%%%%%%

\subsection{Consistency for Model-Parallel Sharding} \label{sec:design:consistency}

\begin{figure}[t]
    \centering
    \includegraphics[width=\linewidth]{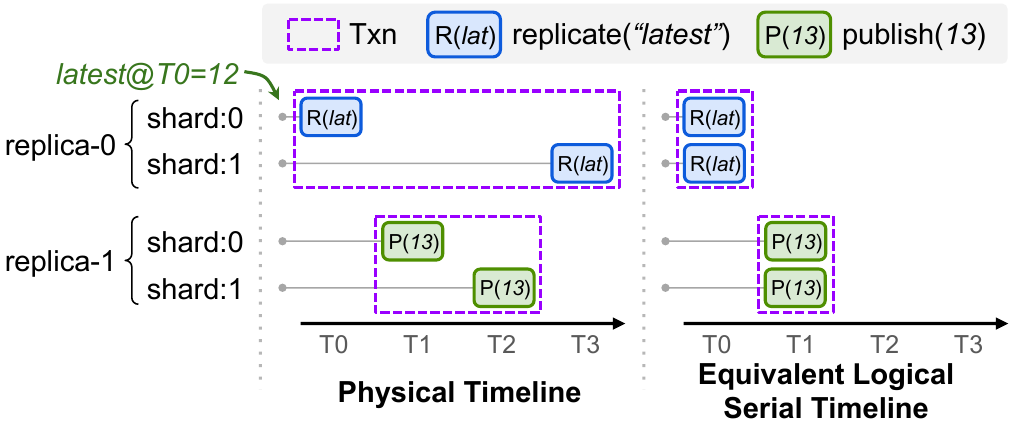}
    \caption{\textbf{Sharding Consistency Example.}
    \it
    Left: the physical request order lets replica-0's shards observe different latest versions: shard:0 sees version 12 at $T_0$, while shard:1 sees version 13 after replica-1's two shards publish it at $T_1$ and $T_2$.
    Without coordination, this difference can cause SPMD control-flow divergence.
    Right: \sysname serializes replica-0's replicate requests as one transaction, so both shards observe version 12.
    }
    \label{fig:txn}
\end{figure}

LLM weights often exceed the memory of a single GPU, so model parallelism is required. 
In this setup, weights are sharded across a group of workers that must execute identical code, following an SPMD (single-program, multiple-data) paradigm.
Divergence in a group risks hanging or corruption.

For correctness, all workers in a model-parallel group should ideally observe a consistent view of the system state, such as whether version $N$ is available or which version is the latest. In practice, however, requests from different workers may reach the server in an interleaved order and observe inconsistent states. Figure~\ref{fig:txn} shows one such example.

\sysname guarantees that all workers in a model-parallel group (\ie, replica) observe the same control-plane state, thereby avoiding inconsistent decisions and SPMD control-flow divergence.
The server implements this guarantee with transactions at replica granularity.
When the first request from a group arrives, the server starts a transaction, evaluates the operation for the entire group, and stores the resulting information (\eg, the latest version) in a transaction-local buffer.
Subsequent requests from that group consume this information.
Transactions are serialized in their start order.

\sysname does not enforce control-plane consistency across replicas.
Each replica polls \sysname independently and may observe a newer version at a different time.
This behavior is acceptable for asynchronous, off-policy RL workloads, which allow rollouts to use slightly stale weights~\cite{AsyncRLHF}.
If cross-replica agreement is required, the RL framework can simply broadcast a message to all workers to pull a specified version from \sysname.

%%%%%%%%%%%%%%%%%%%%%%%%%%%%%%%%%%%%%%%%%%%%%%%%%%%%%%%%%%%%%%%%%%%%%%%%%%%%%%%%

\subsection{Robustness to Churn and Failures} \label{sec:design:fault}

Large-scale training runs on clusters with highly variable reliability. Rollout workers on spot instances can be preempted at any time, and even non-spot nodes occasionally fail at scale.  Our storage abstraction must therefore assume frequent replica churn and provide correctness without manual recovery. \sysname is designed for this failure model: it detects failed replicas, safely re-routes ongoing replication, and integrates cleanly with our retention semantics.

\minititle{General Failure Handling.} 
\sysname considers a client failed if no heartbeat arrives within the timeout window.
Failure is handled at the replica granularity: the server removes the entire failed replica and releases associated resources.

Failures can occur during active replication. If a replica fails while fetching data, the timeout causes the server to invalidate the incomplete replica and release the reference count it acquired. Conversely, if a replica fails while acting as a source, the receiving replica detects the transfer failure and reports it to the server. The server marks the source as failed, preventing future replication from it, and selects an alternate source to recover the transfer. 

\minititle{Retention under Frequent Churn.}
Spot instances change the failure model: failures are not rare events but the norm. To ensure retention with spot instances, when counting the replicas that satisfy a retention rule, \sysname excludes replicas hosted on spot instances.

In the pathological case where the last non-spot replica of a retained version fails, the system can no longer serve that version. In that event, \sysname returns a graceful error indicating that the version is unavailable, and the client can retry later on another version. This behavior is acceptable in RL training because a new version will be trained and published shortly. The consequence of this corner case is a brief delay, not loss of training progress.

\minititle{Reference Server Failure.}
Because the reference server maintains only soft state (\eg, references), recovery is simple and lightweight.
Upon detecting a server failure, a client resets its state to unpublished and connects to a preconfigured backup server.
Critically, this new server does not need to recover any state from its predecessor; it simply waits to be populated by the next round of publishing from trainers.
Before the new server is populated, rollouts continue using the existing weights without interruption.
Eventually, all clients notice the failure and switch to the new server.

A special case arises when the server fails in the middle of a replica transaction, so some shards complete the operation while others observe the failure.
To avoid divergence, the backup server uses a per-replica barrier: it waits for all shards to switch before allowing the replica to proceed.
If some shards do not switch within a timeout, it returns an error to the RL framework, which can then restart this replica.

%%%%%%%%%%%%%%%%%%%%%%%%%%%%%%%%%%%%%%%%%%%%%%%%%%%%%%%%%%%%%%%%%%%%%%%%%%%%%%%%

\subsection{Implementation} \label{sec:design:implementation}

\sysname is implemented with 3.7~K lines of Python and 2.4~K lines of C++ code.
The reference server is implemented as a Ray~\cite{Ray} named actor, instantiated lazily upon the first client call to \api{open()}.
The \sysname transfer engine is built on the Mooncake transfer engine~\cite{Mooncake} for RDMA.

\minititle{Consistency Testing.} 
To validate the server's consistency guarantees, we construct unit tests in which a single test process issues requests on behalf of multiple clients to deterministically simulate arbitrary request interleavings.
Failures are injected by simply halting requests for a given client.
These tests operate solely at the reference-request layer, without involving data transfer, and therefore do not require RDMA NICs or GPUs.
Because all requests originate from a single process, the resulting executions are deterministic and reproducible, greatly simplifying debugging.
This approach to simulated concurrency testing is inspired by FoundationDB~\cite{FoundationDB}, which has demonstrated in production its effectiveness at uncovering subtle concurrency bugs in distributed systems.

\minititle{End-to-End Checksum.} 
To verify end-to-end correctness, we further implement checksums. Upon publishing, the client computes the checksum of each tensor and attaches it to the reference. When another client acquires this reference as a data source, it validates the checksum after transfer.
In addition, these checksums help confirm correct enforcement of the mutability contract without corruption.
Since tensors typically reside on GPUs, checksum computation is fast and can often be overlapped with RDMA transfer.

\section{Evaluation} \label{sec:evaluation}

Our evaluation answers five questions:
\emph{(1) Can \sysname transfer weight tensors efficiently and fully utilize RDMA bandwidth?
(2) Can \sysname scale replication throughput under bursty requests?
(3) Can \sysname mask failure transparently?
(4) Can a single reference server sustain control-plane operations at scale?
(5) Does \sysname uniformly handle a variety of realistic RL workloads?}

We begin with microbenchmarks to evaluate \sysname's data-plane efficiency, scalability under bursts, resilience to failure, and reference server capacity.
We then present three case studies of distinct rollout workloads, focusing on (i) standalone rollouts, (ii) elastic rollouts on spot instances, and (iii) cross-datacenter rollouts, showing how \sysname optimizes transfer with topology, supports dynamic cluster membership, and minimizes worker coordination, all without extra data movement or space.

\minititle{Correctness and Validation.}
We validate the correctness of all experiments using the mechanisms described in \S\ref{sec:design:implementation}. We implemented 1.3~K lines of concurrency tests that cover model-parallel consistency and failure cases. Each experiment runs a separate pass to verify end-to-end checksums on weights, and we trained real models to confirm that loss and accuracy match those from \company production.

\minititle{Hardware Specification.}
Unless otherwise specified, our experiments use machines with 160 CPU cores, 8 NVIDIA Hopper GPUs, 1800~GB of DRAM, 4 Mellanox InfiniBand RDMA NICs (400~Gbps), and 1 VPC NIC (200~Gbps).
The reference server microbenchmark (\S\ref{sec:evaluation:microbench:server}) uses a CPU-only cluster with 32 CPU cores and 32~GB of DRAM per machine.

\subsection{Microbenchmarks} \label{sec:evaluation:microbench}

We use four microbenchmarks to evaluate \sysname's data-plane efficiency, scalability under bursts, failure resilience, and reference server capacity.
Unless otherwise specified, each worker group consists of 8 shards on an 8-GPU machine.

\subsubsection{Data-Plane Bandwidth Efficiency} \label{sec:evaluation:microbench:bandwidth}

In the first microbenchmark, one machine serves as the trainer group and sends tensors, while another serves as the rollout group and receives tensors from the corresponding shards.
We vary the number of tensors per shard to control shard size (50~MB per tensor) and measure the latency required to complete the transfer.
We compare with NCCL~\cite{NCCL}, UCX~\cite{UCX}, and the Ray object store~\cite{Ray}, and show an additional \emph{RDMA ideal} curve that denotes the theoretical number if the RDMA bandwidth is fully saturated (25~GB/s per shard).

\begin{figure}
    \centering
    \begin{subfigure}[b]{0.49\linewidth}
          \includegraphics[width=\linewidth]{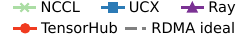}
          \includegraphics[width=\linewidth]{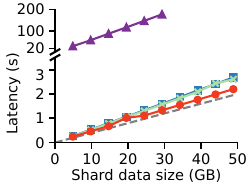}
          \caption{Transfer latency.}
          \label{fig:experiment:microbench:size}
    \end{subfigure}
    \begin{subfigure}[b]{0.49\linewidth}
          \includegraphics[width=\linewidth]{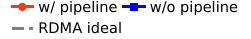}
          \includegraphics[width=\linewidth]{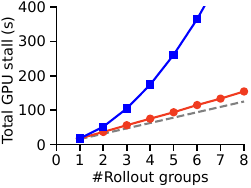}
          \caption{Pipeline replication.}
          \label{fig:experiment:microbench:pipeline}
    \end{subfigure}
    \begin{subfigure}[b]{0.49\linewidth}
          \includegraphics[width=\linewidth]{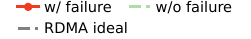}
          \includegraphics[width=\linewidth]{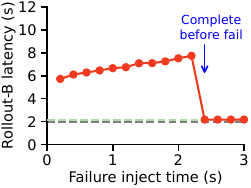}
          \caption{Transfer latency upon failure.}
          \label{fig:experiment:microbench:failure}
    \end{subfigure}
    \begin{subfigure}[b]{0.49\linewidth}
          \includegraphics[width=\linewidth]{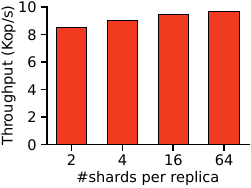}
          \caption{Reference server throughput.}
          \label{fig:experiment:microbench:server}
    \end{subfigure}
    \caption{\textbf{Microbenchmark Results.}}
    \label{fig:experiment:microbench}
\end{figure}

Figure~\ref{fig:experiment:microbench:size} shows the results, where \sysname consistently delivers the lowest latency.
The Ray object store is two orders of magnitude slower than other approaches because it is primarily optimized for CPU memory and TCP, incurring prohibitive costs for GPU--CPU movement and serialization. In addition, Ray crashes at a per-shard size of 35~GB.
Focusing on the three direct transfer approaches, \sysname transfers 50~GB of data in 2.2~s with 22~GB/s throughput (88\% of the theoretical), while NCCL and UCX only achieve 18.8~GB/s and 18.1~GB/s. These results show that \sysname's data-plane implementation is highly efficient and can saturate RDMA bandwidth.

\subsubsection{Scaling with Bursts} \label{sec:evaluation:microbench:pipeline}

In the second microbenchmark, we evaluate \sysname's scalability under bursty request arrival.
We use one machine for a trainer group and 1--8 machines for rollout groups; each shard contains 50~GB of data.
All rollouts simultaneously request replication.

We define the \emph{total GPU stall time} as the sum of stall times over all GPUs involved in the replication.
This metric reflects the aggregated compute resource loss across the cluster.
We report the total GPU stall time for both the default \sysname and the one with pipeline replication disabled, together with the RDMA ideal number for reference.

Figure~\ref{fig:experiment:microbench:pipeline} shows total GPU stall time.  With pipeline replication, the trainer group serves only one rollout group, and the remaining rollouts fetch peer-to-peer. Each shard transfers 50~GB in 2.2~s, independent of replica count. Total stall time scales linearly with the group count and remains close to the RDMA roofline.
Without pipeline replication, all replicas fetch from the trainer, causing bandwidth contention. Stall time grows quadratically with group count.
This scaling gap demonstrates that pipeline replication is necessary for efficient weight dissemination at scale.

\subsubsection{Transparent Failure Masking} \label{sec:evaluation:microbench:failure}

We evaluate the system's resilience to failure with the third microbenchmark.
Since failure at rest does not impact performance, we focus on failure during a transfer.
A resilient system should mask failure without disrupting other healthy workers.

To evaluate this, we construct a pipeline where a trainer group sends data to rollout group A, which then forwards data to rollout group B.
Both rollout groups start simultaneously. Each shard has 50~GB of data. 
We inject a failure into rollout-A sometime after the transfer has begun, and then measure whether rollout-B can complete the transfer without interruption, as well as the total time required.

Figure~\ref{fig:experiment:microbench:failure} presents the results, where rollout-B always completes the transfer.
When the failure occurs immediately after the transfer starts, rollout-B quickly detects it and retries.
The RDMA layer takes a conservative timeout of $\sim$4~s before B confirms A's failure and requests recovery from the server.
The server schedules B to directly read from the trainer to complete the transfer.
For failures injected at later points, B takes longer to complete because retransmission is delayed. If the failure is introduced after 2.2~s, rollout-B has already finished the transfer and is therefore unaffected.
These results demonstrate that \sysname can transparently mask failure with only minor delay.

\subsubsection{Reference Server Capacity} \label{sec:evaluation:microbench:server}

In the last microbenchmark, we measure the sustained throughput of a single reference server.
This benchmark isolates the control plane: clients issue requests but perform no RDMA transfers.
We use one machine to run the server and 64 machines to run a total of 8192 clients.
Each client submits publish, update, and unpublish requests in a closed loop.
We group clients into replicas and vary the number of shards per replica.

Figure~\ref{fig:experiment:microbench:server} shows that a single server sustains 8.5--9.7~Kop/s.
This capacity is sufficient for the control-plane load of RL clusters with thousands of workers.
Throughput does not degrade as replica size increases, showing that model-parallel transactions add little overhead.
Larger replicas reduce the number of replicas and the associated per-replica state, yielding slightly higher throughput.

\subsection{RL Training with Standalone Rollouts} \label{sec:evaluation:basic}
We now evaluate \sysname in realistic RL settings.
We use the \company production framework, which extends veRL's~\cite{HybridFlow} co-located architecture with standalone rollouts.
The framework relies on a Ray driver to orchestrate weight transfer across workers, which is a common and representative design of RL frameworks~\cite{HybridFlow,AReaL,OpenRLHF}.

\begin{table}
    \centering\small
    \begin{tabular}{ccccc}
    \hline\hline
        \textbf{Model size} & 9B & 36B & 260B & \emph{mocked} 1T \\
    \hline
        \textbf{\#shards} & 2 & 4 & 8 & 16 \\
        \textbf{Shard size (GB)} & 10 & 19 & 34 & 66 \\
        \textbf{Trainer \#GPUs} & 16 & 16 & 64 & 768 \\
        \textbf{Standalone \#GPUs} & 8 & 8 & 16 & 256 \\
    \hline\hline
    \end{tabular}
    \caption{\textbf{Training Workload Parameters.} 
    }
    \label{tab:experiments:basic:model}
\end{table}

\minititle{Baseline: NCCL and UCX.}
We compare against two weight transfer systems used in the \company production stack: one based on NCCL~\cite{NCCL} and another based on UCX~\cite{UCX}. 
As discussed in \S\ref{sec:background:existing}, the communication-based weight transfer systems require the RL framework to coordinate workers.
In practice, it is implemented as an RPC broadcast that interrupts all workers for a global weight transfer stage.
More fine-grained coordination is theoretically possible, but it requires a major refactor of the framework.
Though we are unable to compare with every possible implementation, we again include a roofline curve for reference (labeled ``RDMA ideal''), denoting an ideal case with zero coordination cost or contention, bottlenecked only by the rollouts' RDMA bandwidth.
Performance close to this reference implies near-optimal results.
We did not include a comparison with other storage-based approaches because we have shown in \S\ref{sec:evaluation:microbench} that they are orders of magnitude slower.

\begin{figure}[t]
    \centering
    \includegraphics[width=\linewidth]{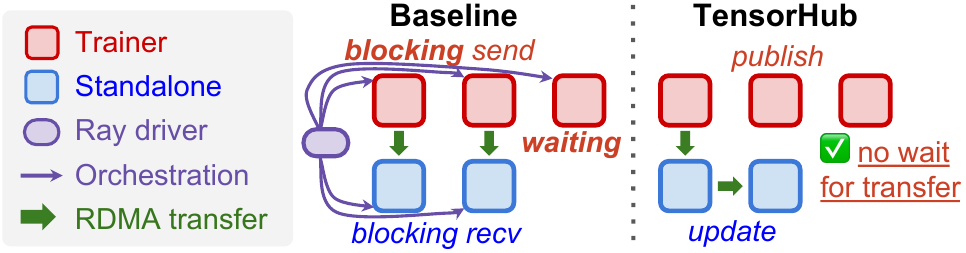}
    \caption{\textbf{Weight Transfer Flows with Standalone.}
    \it
    \sysname does not require the Ray driver to orchestrate weight transfer.
    Each standalone rollout pulls weights on demand.
    }
    \label{fig:experiment:basic:workflow}
\end{figure}

\begin{figure}[t]
    \centering
    \includegraphics[width=\linewidth]{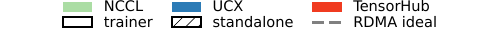}
    \includegraphics[width=\linewidth]{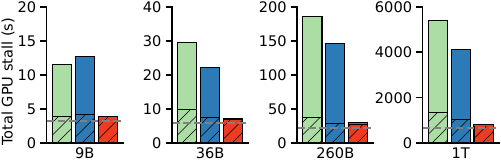}
    \caption{\textbf{Standalone Rollout Results.}
    \it
    Ideally, only standalone rollouts need to stall due to their dependence on weights; trainers can proceed without waiting.
    \sysname not only eliminates trainer stalls, but also keeps standalone stall time consistently lower.
    }
    \label{fig:experiment:basic}
\end{figure}

\minititle{Engineering Effort for Integration.}
NCCL requires $\sim$450 lines of code changes to the \company production RL framework, mostly for coordination.
UCX, with more low-level control, requires $\sim$1200 lines, trading simplicity for flexibility.
\sysname requires only $\sim$40 lines of change (similar to the example in Figure~\ref{fig:code}).
This highlights that \sysname offers a simple and user-friendly interface.

\minititle{Workloads.}
We train four models for evaluation (Table~\ref{tab:experiments:basic:model}), covering various sharding and GPU count setups.
The shard count is the number of GPUs in one model-parallel replica, and the total GPU counts determine the number of trainer and standalone replicas.
For example, the 260B configuration uses 8-shard replicas, yielding 8 trainer replicas and 2 standalone replicas.
The 9B, 36B, and 260B models are initialized from pretrained weights.
To stress the system at an extreme scale, we construct a 1T-parameter model with mocked weight data by duplicating the 260B-parameter model's layer weights four times: training such a huge model requires 768~GPUs for trainers and 256~GPUs for standalone rollouts (1024~GPUs in total).

\minititle{Results.}
Figure~\ref{fig:experiment:basic:workflow} compares the workflows of the baselines and \sysname, and Figure~\ref{fig:experiment:basic} shows that \sysname reduces total GPU stall time by up to 6.7$\times$ compared with NCCL and by 5.2$\times$ compared with UCX.
These reductions stem from \sysname's minimal-coordination design.
First, \api{publish} is a lightweight reference-passing request that does not wait for actual data transfer; trainers immediately resume for their own co-located rollout tasks. 
In contrast, NCCL and UCX require coordination by the framework's Ray driver, and the driver does not move forward to its next stage until the weight transfer stage is completed.

Second, for large models, stragglers are amplified at scale: the weight transfer stage completes only when the slowest worker finishes. 
For the 1T model, transferring 66~GB should ideally take 2.6~s at full bandwidth, but NCCL and UCX stall all 1024~GPUs for 5.3~s and 4.0~s, respectively. 
\sysname localizes the stall: mean latency is 3.1~s, and only the standalones (256~GPUs) need to wait.
This demonstrates the benefit of decoupling trainer and rollout control flow.

\subsection{Elastic Rollouts on Spot Instances} \label{sec:evaluation:elastic}

In the second case study, we evaluate \sysname when rollout workers run on a hybrid of stable GPUs (for standalone) and preemptible spot GPUs (for elastic). 
In the \company production environment, an autoscaler monitors training progress and instantiates new elastic rollouts when rollout tasks are backlogged and spare spot GPUs are available. 
When their GPU resources are preempted, elastic rollouts are killed immediately without a grace period.

\minititle{Baseline.}
Only UCX-based weight transfer supports elastic rollout, because NCCL assumes static membership. Supporting elastic rollouts took substantial engineering effort for the \company production UCX-based implementation. When a new elastic worker appears, trainers may be mid-training and unable to serve weights, so standalone rollouts must also act as data senders. This introduces a trainer-to-standalone-to-elastic transfer topology.

\begin{figure}[t]
    \centering
    \includegraphics[width=\linewidth]{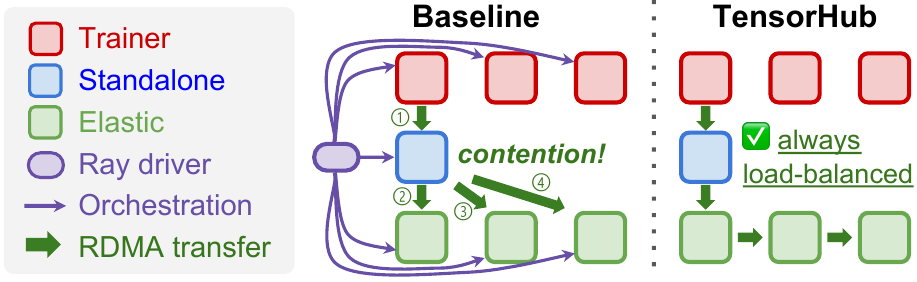}
    \caption{\textbf{Weight Transfer Flows with Standalone and Elastic.}
    \it
    Note that \sysname only shows one possible data flow; an elastic rollout can also fetch weights from a trainer.
    }
    \label{fig:experiment:elastic:workflow}
\end{figure}

\begin{figure}[t]
    \centering
        \includegraphics[width=\linewidth]{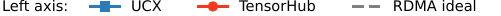} \\
        \vspace{-4pt}
        \includegraphics[width=\linewidth]{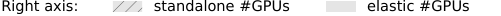} \\
    \begin{subfigure}[b]{0.66\linewidth}
        \includegraphics[width=\linewidth]{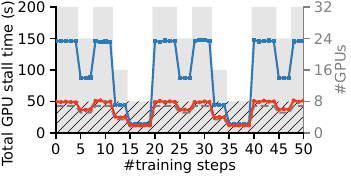}
        \caption{\textbf{Rollout GPU stall time over steps.}\\
        \it
        The curves (left axis) show stall time; the shading (right axis) shows GPU counts.
        Note that only standalone and elastic stall time is shown; UCX's additional trainer stall is omitted.
        }
        \label{fig:experiment:elastic:timeline}
    \end{subfigure}
    \hfill
    \begin{subfigure}[b]{0.31\linewidth}
        \includegraphics[width=\linewidth]{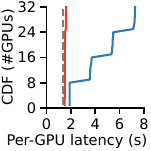}
        \caption{\it Latency CDF at step~10; 8 GPUs for the standalone rollouts and 24 for the elastic rollouts.}
        \label{fig:experiment:elastic:cdf}
    \end{subfigure}
    \caption{\textbf{Elastic Rollout Results.}
    }
    \label{fig:experiment:elastic}
\end{figure}

\minititle{Workloads.}
We use the 260B model described in Table~\ref{tab:experiments:basic:model}, but with one standalone machine (8~GPUs) and up to three elastic machines (24~GPUs). To keep experiments reproducible, we intercept the autoscaler and deterministically issue scale-up and scale-down events. Randomly chosen elastic replicas are terminated during scale-down.

\minititle{Results.}
Figure~\ref{fig:experiment:elastic:workflow} compares the workflows, and Figure~\ref{fig:experiment:elastic:timeline} shows total rollout stall time over training steps with GPU count varied.
The stall time covers both standalone and elastic rollout GPUs; UCX's additional trainer stall is omitted.

In the UCX baseline, elastic rollouts must wait for standalone rollouts to pull weights from trainers first, introducing a delay. When elastic workers outnumber standalones, bandwidth contention further amplifies tail latency. Figure~\ref{fig:experiment:elastic:cdf} shows per-GPU stall time at step 10: UCX produces a stair-shaped CDF, with the last batch of elastic GPUs waiting up to 7.2~s for weights.

With \sysname, the reference server balances load with topology awareness at runtime, so elastic rollouts can fetch weights directly from trainers, standalones, or peer elastic rollouts, and pipeline replication spreads load across all available replicas. As a result, per-GPU stall time remains near-constant ($\sim$1.5~s) independent of the elastic GPU count, approaching the RDMA ideal.
\sysname reduces transfer latency by up to 4.8$\times$ compared with UCX.
This result demonstrates that \sysname naturally accommodates dynamic membership and runtime topology.

\subsection{Cross-Datacenter RL Training} \label{sec:evaluation:cross_dc}

\begin{figure}[t]
    \centering
    \includegraphics[width=\linewidth]{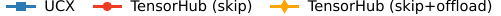}
    \includegraphics[width=\linewidth]{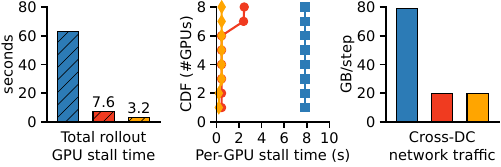}
    \caption{\textbf{Cross-Datacenter Rollout Results.}
    \it
    Note that the left figure shows only the rollout stall time; UCX incurs additional trainer stall time that is omitted here.
    }
    \label{fig:experiment:cross_dc}
\end{figure}

The final case study evaluates \sysname in a multi-datacenter deployment. This setting reflects production scenarios where spare inference resources are available in a separate datacenter, and utilizing them requires weight transfers over slower inter-datacenter networks.

\minititle{Baseline.}  We use a UCX-based implementation with TCP transport as the baseline. The transfer topology is similar to that of standalone rollouts (Figure~\ref{fig:experiment:basic:workflow}), except the data path is now cross-datacenter over TCP.

\minititle{Workloads.}
We use the 9B model from Table~\ref{tab:experiments:basic:model}, with eight 2-shard trainer replicas (16~GPUs) and four 2-shard rollout replicas (8~GPUs). Trainers are located in a datacenter with training-optimized Hopper GPUs; rollouts run in a second datacenter with inference-optimized Hopper GPUs. The two datacenters are connected only via the 200~Gbps VPC NIC (one per machine) with TCP.

\minititle{Results.}
Figure~\ref{fig:experiment:cross_dc} reports total stall time across rollout GPUs, per-GPU latency distribution, and total cross-datacenter TCP traffic. \sysname reduces total rollout GPU stall time by up to 19$\times$ compared with UCX. With the UCX baseline, every rollout replica fetches weights directly over TCP. Since all flows contend for the single VPC NIC per machine, stall time is dominated by cross-datacenter bandwidth: each GPU waits 7.8~s for weights.
Such waiting is costly, given that an end-to-end training step takes only 68~s.

With \sysname and smart skipping, only the first rollout replica (2~GPUs) waits for cross-datacenter TCP transfer (2.5~s), and the contention on the VPC NIC is reduced.
Other rollout replicas can fetch data via high-throughput RDMA with merely 0.45~s latency.  The result is a latency distribution with a single long tail for the seeding replica and near-ideal RDMA latency for all others.
The end-to-end training step time is reduced to 60~s, 13\% faster than UCX.

With offload seeding enabled, \sysname further hides the cross-datacenter transfer latency by using an offload replica.
The seeding replica stalls only for the 0.21~s PCIe transfer.
In summary, \sysname exploits datacenter locality and achieves near-local performance after a single cross-datacenter seeding transfer.

\section{Experience and Lessons} \label{sec:experience}

We share our experience and lessons learned when operating the weight transfer systems in production.

\minititle{Clean Interfaces Avoid Deadlocks.}
Weight transfer introduces synchronization across worker groups and hence is vulnerable to deadlock.
In earlier NCCL/UCX-based weight transfer systems, we observed a series of related bugs.
For example, one deadlock arose from a circular dependency between the Ray driver and workers: the driver asked all rollouts to stop; a standalone rollout acquired a write lock on its weights, while a bootstrapping elastic rollout blocked trying to read those weights from the standalone and could not respond to the driver; the driver then waited indefinitely.
Such bugs arise because weight transfer creates a large critical section, and each worker runs customized logic to enter it, potentially introducing more dependencies.

In contrast, \sysname APIs avoid this dependency pattern.
\api{publish} is non-blocking; \api{update} returns once the transfer is complete; and \api{unpublish} waits only for ongoing readers to drain.
Each API has fixed blocking behavior to minimize the critical section.
No external, framework-specific coordination logic enters the critical section, preventing circular dependencies by design.

\minititle{Decoupling Isolates Delays.}
\sysname's decoupled interfaces also help to isolate the impact of suboptimal implementations.
For example, in earlier systems we noticed cases where a trainer incorrectly ran manual garbage collection or expensive weight preprocessing in the middle of a weight transfer stage, causing unnecessary stalls for rollouts.
\sysname APIs block a rollout only for its transfer time, preventing unrelated preprocessing from ever entering the critical section.

\section{Related Work} \label{sec:related}

\minititle{Reinforcement Learning for LLMs.}
Many frameworks~\cite{OpenRLHF,HybridFlow,RLHFuse} have been proposed to accelerate RL training. veRL~\cite{HybridFlow} combines single-controller and multi-controller paradigms to drive stage execution efficiently. RLHFuse~\cite{RLHFuse} further improves training compute efficiency with stage fusion. RollPacker~\cite{RollPacker} optimizes the long-tail decoding in rollout, while SpecActor~\cite{SpecActor} introduces decoupled and Best-of-N speculation to accelerate the rollout phase, ensuring on-policy model training. RLBoost~\cite{RLBoost} harvests preemptible GPUs to speed up the rollout phase with resource elasticity.
Recent work also explores off-policy approaches for training acceleration.
AsyncRLHF~\cite{AsyncRLHF}, AReaL~\cite{AReaL}, StreamRL~\cite{StreamRL}, and Laminar~\cite{Laminar} introduce asynchronous RL post-training with tailored optimizations to increase job throughput.
TLT~\cite{TLT} trains a draft model in parallel for speculation when RL model training is performed; thus, the draft model also requires weight synchronization.

These frameworks optimize RL scheduling, which increasingly requires efficient weight transfer; they can thus benefit from \sysname's clean weight transfer abstraction.

\minititle{Storage Systems for Machine Learning.}
Prior studies have built storage systems for machine learning workloads.
Parameter Server~\cite{ParameterServer} stores model weights across a cluster of servers.
Ray~\cite{Ray} provides an object store for RL, but is primarily optimized for CPU workloads.
Hoplite~\cite{Hoplite} extends Ray with fault-tolerant collective primitives.
Unlike \sysname, these systems require a system-owned copy of model weights, which is prohibitively expensive for LLM weights at TB scale.

Checkpointing systems are another class of storage for weights.
Unlike RL weight transfer, they retain model states for recovery and are less latency-sensitive.
GeminiFS~\cite{GeminiFS} accelerates checkpointing and activation offloading by optimizing the metadata through a POSIX-like interface.
ByteCheckpoint~\cite{ByteCheckpoint} is designed for efficient model weight storage and supports dynamic resharding across different training parallelism configurations.
Gemini~\cite{Gemini} stores in-memory checkpoint replicas for faster failure recovery.

\minititle{Peer-to-Peer Data Dissemination.}
The idea of serving data requests from peer clients instead of servers dates back to cooperative caching~\cite{Cooperative_Caching} and xFS~\cite{xFS}.
However, retaining additional copies on clients is too costly for LLM weights.
\sysname avoids this space overhead by borrowing the application-owned copies via a mutability contract and a retention protocol.
Ekko~\cite{Ekko} disseminates recommender-model updates peer-to-peer across WANs, exploiting their sparsity and locality with eventual consistency.
In contrast, \sysname focuses on transferring dense LLM weights on the performance-critical path of RL workloads.

\section{Conclusion} \label{sec:conclusion}

We present Reference-Oriented Storage (ROS), a new approach to weight transfer in large-scale LLM reinforcement learning systems. 
By eliminating explicit data ownership and instead leveraging the inherent redundancy and immutability of model weights, ROS reconciles a long-standing trade-off between flexibility and efficiency in RL weight transfer systems.
Building on this abstraction, \sysname enables topology-aware, low-overhead weight transfer with minimal coordination, while preserving model-parallel consistency and fault tolerance in dynamic environments.
Our evaluation shows that \sysname can saturate network bandwidth and deliver near-optimal end-to-end performance.
\sysname has been deployed in \company production for cutting-edge RL training.

\begin{acks}
We thank our shepherd, Juncheng Yang, and the anonymous reviewers for their helpful comments.
We are also grateful to Jing Liu, Shivaram Venkataraman, Ming Liu, Mai Zheng, and Yuhong Zhong for their constructive feedback, and to Siyuan Chai for assistance with the reference server experiment.
\end{acks}

\bibliographystyle{ACM-Reference-Format}
\bibliography{reference}

\end{document}